\documentclass[a4paper]{jpconf}
\usepackage{graphicx}
\begin{document}

\title {Casimir pressure on a thin metal slab}

\author { M. S. Toma\v{s}$^1$ and Z. Lenac$^2$ }

\address {$^1$Rudjer Bo\v{s}kovi\'{c} Institute , P. O. Box 180,
10002 Zagreb, Croatia }
\address {$^2$Department of Physics, University of Rijeka, 51000 Rijeka, Croatia }

\ead{tomas@thphys.irb.hr}

\begin{abstract}
We consider the vacuum-field pressure on boundaries of a metal
slab in the middle of a cavity with perfectly reflecting mirrors
adopting the plasma model for the metal and paying special
attention to the surface plasmon polariton contribution to the
pressure. We demonstrate that, with increasing cavity length, the
pressure on a thin ($d\ll\lambda_P$) slab in this system decreases
from the Casimir pressure $F_C=-\pi^2\hbar c/240 d^4$ at zero
slab-mirror distances to the non-retarded force per unit area
$F_{\rm nr}=1.19 (d/\lambda_P)F_C$ in the case of an isolated
slab. In the first case the pressure is entirely due to the
photonic modes propagating through the metal whereas in the second
case it is entirely due to the (nonretarded) surface plasmon modes
supported by the free-standing thin slab. In either case the
pressure decreases with the slab thickness. These considerations
indicate that the vacuum-field pressure on a thin metal layer (and
its modal structure) can be in a symmetric cavity significantly
influenced when changing the cavity length.
\end{abstract}

\section{Introduction}
The Casimir effect most commonly refers to the existence of forces
between neutral macroscopic bodies due to the change of the
zero-point energy of the electromagnetic field in confined space
\cite{Mill,BMM,Milt,Lamo}. Soon upon its prediction for two
perfectly conducting plates in vacuum \cite{Cas}, the theory of the
Casimir effect was extended to more realistic systems consisting of
two dielectrics separated by a vacuum gap \cite{Lif} and by a gap
filled by a medium \cite{Dzy,Abr,Schw} and, more recently, to
systems involving more layers \cite{Zhou} including general
dielectric and magnetodielectric multilayers
\cite{Tom,Raab,Tom2,Raab2}. Evidently, when the gap between two
stacks of layers is filled by a medium, as in multilayers, the
vacuum-field force (per unit area) on the stacks can also be
regarded as the pressure on the medium between them. Moreover, since
the vacuum-field fluctuations are always present, the pressure on
the medium persists even in the absence of other layers, that is,
even in the case of a free standing single material slab.
Consequently, as pointed out recently \cite{BeCa}, in addition to
the traditional Casimir force due to the presence of other layers,
every material layer in a multilayered systems experiences a
vacuum-field pressure on its boundaries. Changes caused by this
pressure (e.g. change of the layer thickness) can be taken as an
alternative signature of the Casimir effect and it is therefore of
fundamental interest to explore it in more details. From the
practical point of view, however, particularly interesting systems
in this respect are those involving metallic plates as metal
components are often met in micromechanical (MEMS) and
nanomechanical (NEMS) devices \cite{Macl,GuDe}.

Vacuum-field pressure on a metal slab has already been addressed
(to some extent) by Dzyaloshinskii {\it et al}. \cite{Dzy} when
discussing the Casimir force between two dielectric media
separated by a metal layer. The same system has also recently been
considered by Imry \cite{Imry} who pointed out strong dependence
of the zero-point radiation pressure on a metal film on properties
of the surrounding media. Very recently, Benassi and Calandra
\cite{BeCa} used the Lifshitz formula to calculate the pressure on
a metal slab and explore its dependence on the properties of the
slab as well as on the distance and properties of a nearby
(metallic) substrate. By combining a Lifshitz-like formula and the
mode summation method, in our previous work \cite{LeTo} we have
explored the effect of surrounding media as well as of a cavity on
the pressure on a metal slab paying special attention to the
contribution of the surface polariton (SP) modes to the pressure.
In the present work, we consider in more details the effect of a
cavity on the pressure on surfaces of a metal slab in its center
and derive several new results concerning this pressure. Since the
ordinary Casimir force on the slab vanishes in symmetric
configurations, consideration of this system is a very convenient
way to explore the pressure on the slab surfaces. As in
aforementioned works, we adopt free-electron (plasma) model to
describe the metal and, for simplicity, assume perfectly
reflecting cavity mirrors. Accordingly, our aim here is to
establish trends of the pressure with the system parameters and
calculate its limiting values rather than to discuss it for a
realistic system.

The paper is organized as follows. In Section II, we briefly adapt
the theory of the ordinary Casimir force on a slab in a cavity
\cite{Tom,Raab2, Elli} to include also the vacuum-field pressure
on the slab surfaces and derive a Lifshitz-like formula for this
pressure (see also \cite{BeCa}). In Sections III-V, we use this
formula to discuss the pressure on surfaces of a free standing
metal slab, of a metal layer between perfect mirrors and of a
metal slab in an ideal cavity, respectively. Our conclusions are
summarized in Section VI.

\section{Preliminaries}

Consider a dielectric slab inserted in a planar cavity, as
depicted in Fig. 1. The total force (per unit area) acting on the
slab consists of the pressure $F=f_s$ on the slab boundaries and
the slab-mirror interaction force $F'=f_2-f_1$ \cite{BeCa}, where
\begin{figure}[h]
\begin{center}
\includegraphics[width=20pc]{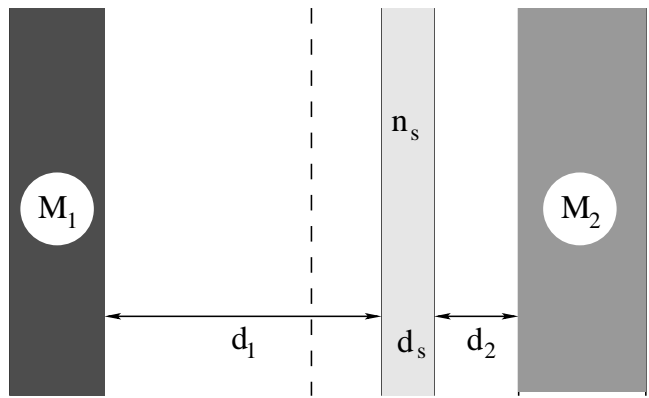}
 \end{center}
 \caption{\label{sys} System considered schematically. }
\end{figure}
according to the theory of the Casimir force in multilayers
\cite{Tom}
\begin{equation}
\label{fj}
f_j=-\frac{\hbar}{2\pi^2}\int_0^\infty d\xi
\int^\infty_0 dkk\kappa_j\sum_{q=p,s} \frac{1-D_{qj}(i\xi,k)}
{D_{qj}(i\xi,k)},\;\;\;D_{qj}(i\xi,k)=1-r^q_{j-}r^q_{j+}e^{-2\kappa_jd_j}.
\end{equation}
Here $\kappa_j(i\xi,k)=\sqrt{\varepsilon_j(i\xi)\xi^2/c^2+k^2}$ is
the perpendicular wave vector at the imaginary frequency in the
$j$th layer and $r^q_{j\pm}(i\xi,k)$ are the reflection
coefficients of the right and left stack of layers bounding the
layer.

Considering first the pressure on the slab, we have
\begin{equation}
\label{fsg}
\fl F=-\frac{\hbar}{2\pi^2}\int_0^\infty d\xi
\int^\infty_0
dkk\kappa_s\sum_{q=p,s}\frac{r^q_{s-}r^q_{s+}e^{-2\kappa_s
d_s}}{1-r^q_{s-}r^q_{s+}e^{-2\kappa_s d_s}},
\end{equation}
where
\begin{equation}
\label{rq} r^q_{s-(+)}(i\xi,k)=\frac{-\rho^q+R^q_{1(2)}e^{-2\kappa
d_{1(2)}}} {1-\rho^qR^q_{1(2)}e^{-2\kappa d_{1(2)}}}
\end{equation}
are reflection coefficients for the waves reflected within the
slab. Here
$\kappa(i\xi,k)\equiv\kappa_1=\kappa_2=\sqrt{\xi^2/c^2+k^2}$ is
the perpendicular wave vector in the cavity,
\begin{equation}
\label{rho}
 \rho^p(i\xi,k)=\frac{\varepsilon_s\kappa
-\kappa_s}{\varepsilon_s\kappa +\kappa_s },\hspace{1 cm}
\rho^s(i\xi,k)=\frac{\kappa -\kappa_s}{\kappa +\kappa_s },
\end{equation}
are the {\it vacuum-slab} reflection coefficients and
$R^q_{1(2)}(i\xi,k)$ are those of the mirrors. According to Eqs.
(\ref{fsg}) and (\ref{rq}), $f_s$ can be rewritten as
\begin{equation}
\fl F=-\frac{\hbar}{2\pi^2}\int_0^\infty d\xi \int^\infty_0
dkk\kappa_s\sum_{q=p,s}\frac{(\rho^q-R^q_1e^{-2\kappa
d_1})(\rho^q-R^q_2e^{-2\kappa d_2})e^{-2\kappa_s
d_s}}{\tilde{D}_{qs}}, \label{fs}
\end{equation}
where
\begin{eqnarray}
\fl \tilde{D}_{qs}(i\xi,k)&=&1-{\rho^q}^2e^{-2\kappa_s
d_s}-\rho^q(1-e^{-2\kappa_s d_s})(R^q_1e^{-2\kappa
d_1}+R^q_2e^{-2\kappa d_2})\nonumber\\
&&+({\rho^q}^2-e^{-2\kappa_s d_s})R^q_1R^q_2e^{-2\kappa(d_1+d_2)}.
\label{tD}
\end{eqnarray}

The traditional force per unit area on the slab due to the
presence of the mirrors $F'=f_2-f_1$ is found similarly using
\cite{Tom}
\begin{equation}
r^q_{1-(2+)}(i\xi,k)=R^q_{1(2)},\hspace{1 cm}
r^q_{1+(2-)}(i\xi,k)=r^q+\frac{{t^q}^2R^q_{2(1)}e^{-2\kappa
d_{2(1)}}} {1-r^qR^q_{2(1)}e^{-2\kappa d_{2(1)}}},
\end{equation}
where $r^q(i\xi,k)$ and $t^q(i\xi,k)$
are the Fresnel coefficients for the (whole) slab. We find that
$F'$ is given by Eq. (\ref{fs}) provided that the nominators in
that equation are replaced by \cite{Tom}
\begin{equation}
\rho^q(1-e^{-2\kappa_s d_s})(R^q_2e^{-2\kappa
d_2}-R^q_1e^{-2\kappa d_1})
\end{equation}
and that $\kappa_s\rightarrow\kappa$ in front of the sum. Clearly,
in contrast to $F$, the pressure $F'$ vanishes when the slab is in
the center ($d_1=d_2$) of a symmetric ($R^q_1=R^q_2$) cavity.

Frequencies $\omega^q_n(k)$ of SP and other bound modes supported
by the system are given as solutions of
\begin{equation}
D_{qs}(\omega,k)\equiv
1-r^q_{s-}r^q_{s+}e^{-2\alpha_sd_s}=0,\;\;\;
\alpha_s(\omega,k)\equiv\kappa_s(-i\omega,k)=\sqrt{k^2-\varepsilon_s\omega^2/c^2},
\end{equation}
[or, equivalently, $\tilde{D}_{qs}(\omega,k)=0$] in the
corresponding part of the $(\omega,k)$-plane. Separate
contribution of each SP mode to the Casimir force can therefore be
obtained by using the real frequency counterpart of Eq. (\ref{fs})
and calculating the corresponding residuum. Alternatively, since
this equation is (for a lossless system) compatible with the
standard definition of the Casimir energy (with respect to the
slab)
\begin{equation}
E_s = A \int \frac{d^2{\bf k}}{(2 \pi)^2} \sum_{q=p,s} \sum_n
\frac{\hbar}{2} \left[ \omega_{n}^{q}(k,d_s)  -
\omega_{n}^{q}(k,d_s \rightarrow \infty) \right],
 \label{EC}
\end{equation}
we may calculate the surface contribution to the pressure using
directly
\begin{equation}
\label{FSP} F_S= -\frac{1}{A}\frac{ \partial E^{\rm SP}_s
}{\partial d_s }=-\frac{\hbar}{4\pi}\int_0^\infty dkk\sum_\sigma
\frac{\partial\omega_\sigma(k)}{\partial d_s},
\end{equation}
where $\sigma$ enumerates SP modes. In the following, we use this
method to perform a modal analysis of the pressure on a thin metal
slab in a symmetric cavity. Clearly, in such a configuration,
frequencies of SP (and other) modes are found as solutions of
\begin{equation}
\label{DR}
 r^q_s(\omega,k)e^{-\alpha_s d_s}=\pm 1,\;\;\;
r^q_s(\omega,k)=\frac{-\rho^q+R^qe^{-2\alpha d_s}}
{1-\rho^qR^qe^{-2\alpha d_s}},\;\;\;
\alpha(\omega,k)=\sqrt{k^2-\omega^2/c^2},
\end{equation}
and the modes are further characterized by an index $\nu=\pm$
describing their symmetry with respect to the central plane of the
system.

\section{Free-standing metal slab}
Let us start with the case of a free-standing metallic slab.
Adopting the electron plasma model, the slab is described by the
dielectric function
\begin{equation}
\label{eps}
\varepsilon_s(i\xi)=1+\frac{\omega^2_P}{\xi^2},
\end{equation}
where $\omega_P$ is the metallic plasma frequency. Letting
$R^q_{1(2)}=0$ Eq. (\ref{fs}), the pressure on the slab is then
given by
\begin{equation}
F=-\frac{\hbar}{2\pi^2}\int_0^\infty d\xi \int^\infty_0
dkk\kappa_s\sum_{q=p,s}\frac{[\rho^q(i\xi,k)]^2e^{-2\kappa_s
d_s}}{1-[\rho^q(i\xi,k)]^2e^{-2\kappa_s d_s}}, \label{ff}
\end{equation}
with $\rho^q(i\xi,k)$ given by Eqs. (\ref{rho}) and (\ref{eps}).
Note that this formula differs from the corresponding formula for
the standard Casimir force (per unit area) between two metal
half-spaces only in the (explicit) presence of $\kappa_s$ instead
of $\kappa$.

For a thin, $d_s\ll c/\omega_p$, slab the main contribution to $F$
comes from large wave vectors. Accordingly, we may let
$\kappa_s\simeq \kappa\simeq k$ and consequently
$\rho^p\simeq\rho^p_{\rm nr}=(\varepsilon_s-1)/(\varepsilon_s-1)$
and $\rho^s\simeq\rho^s_{\rm nr}=0$ in Eq. (\ref{ff}). Thus, the
pressure on the slab is in this nonretarded (quasistatic)
approximation equal to the pressure on two identical semi-infinite
metals separated by a thin vacuum gap. Introducing
$x=\xi/\omega_P$ and $t=2kd$ as the integration variables, in this
way we obtain
\begin{equation}
\label{fnr} F_{\rm nr}=-\frac{\hbar\omega_P}{16\pi^2d^3}
\int_0^\infty dx \int^\infty_0dtt^2\frac{e^{-t}}{(2x^2+1)^2
-e^{-t}}=-0.00781 \frac{\hbar\omega_P}{d^3}.
\end{equation}

It is easy to see that $F_{\rm nr}$ is entirely due to the two
surface plasmon modes supported by the slab (or two semi-infinite
free-electron metals) \cite{Imry,KNS,GILR,Henk,Bord}. Indeed, from
the nonretarded limit of Eq. (\ref{DR})
\[\frac{\varepsilon_s(\omega)-1}{\varepsilon_s(\omega)+1}e^{-kd_s}=\mp
1\] we find familiar frequencies of surface plasmons
\begin{equation}
\omega_{\pm}(k)=\frac{\omega_P}{\sqrt{2}}\sqrt{1\pm\e^{- k d_s}},
\end{equation}
so that the integrand in Eq. (\ref{FSP}) becomes:
\[-k \; \frac{ \partial \omega_{\pm} (k) }{ \partial d_s } =
 \pm \frac{\omega_P}{2\sqrt{2}} \frac{k^2e^{- k d_s}}{\sqrt{1\pm e^{- k d_s}}}.
\]
Using this in Eq. (\ref{FSP}), the total (nonretarded)
contribution to the pressure of the two plasmon modes is easily
calculated to be equal to $F_{\rm nr}$ given above as a result of
compensation between the pressing ($F^{\rm SP}_-= 7.83 F_{\rm
nr}$) and relaxing ($F^{\rm SP}_+= - 6.83F_{\rm nr}$)
contributions from the $\omega_-$ and $\omega_+$ mode,
respectively.

To estimate the pressure on a thick slab, we proceed in the standard
way \cite{Lif,Dzy} and introduce in Eq. (\ref{ff}) the variable $p$
instead of $k$ by letting $\kappa_s=\sqrt{\varepsilon_s(i\xi)}\xi
p/c$. With $x=\xi/\omega_P$, this gives
\begin{equation}
F\simeq-\frac{\hbar ck_P^4}{2\pi^2}\int_0^\infty dx(1+x^2)^{3/2}
\int_1^\infty dpp^2\sum_{q=p,s}
\frac{[\rho^q(ix\omega_P,p)]^2e^{-2p\sqrt{1+x^2}k_P d_s}}
{1-[\rho^q(ix\omega_P,p)]^2e^{-2p\sqrt{1+x^2}k_P d_s}},
\end{equation}
where $k_P=\omega_P/c=2\pi/\lambda_P$ and
\begin{equation}
\rho^p(ix\omega_P,p)=\frac{(1+x^2)s-x^2p}
{(1+x^2)s+x^2p},\hspace{0.5 cm}
\rho^s(ix\omega_P,p)=\frac{s-p}{s+p},\hspace{0.5 cm}
s=\sqrt{p^2-1+\frac{x^2}{1+x^2}}.
\end{equation}
 Since for
$k_Pd_s\gg 1$ the main contribution to the integral comes from
small-$x$ region, we can let $x=0$ everywhere except in the
exponents where we use $\sqrt{1+x^2}\simeq 1+x^2/2$. Retaining only
the leading terms, in this way we obtain
\begin{eqnarray}
\label{fsf} F&\simeq& -\frac{\hbar ck_P^4}{4\pi\sqrt{\pi
k_Pd_s}}\int_1^\infty
dpp^{3/2}e^{-2pk_pd_s}\left[1+(\frac{\sqrt{p^2-1}-p}
{\sqrt{p^2-1}+p})^2\right]\nonumber\\
&\simeq& -\frac{\hbar ck_P^4}{4}\frac{e^{-2k_pd_s}}
{\sqrt{(\pi k_Pd_s)^3}},\hspace{1 cm}k_Pd_s\gg 1,
\end{eqnarray}
where the final result follows upon a partial integration.
Accordingly, as already noted by Dzyaloshinskii {\it et al}
\cite{Dzy}, in the plasma model for the metal the pressure on the
slab surfaces exponentially decreases with $d_s$.

\begin{figure}[htb]
\begin{minipage}{18pc}
\includegraphics[width=18pc]{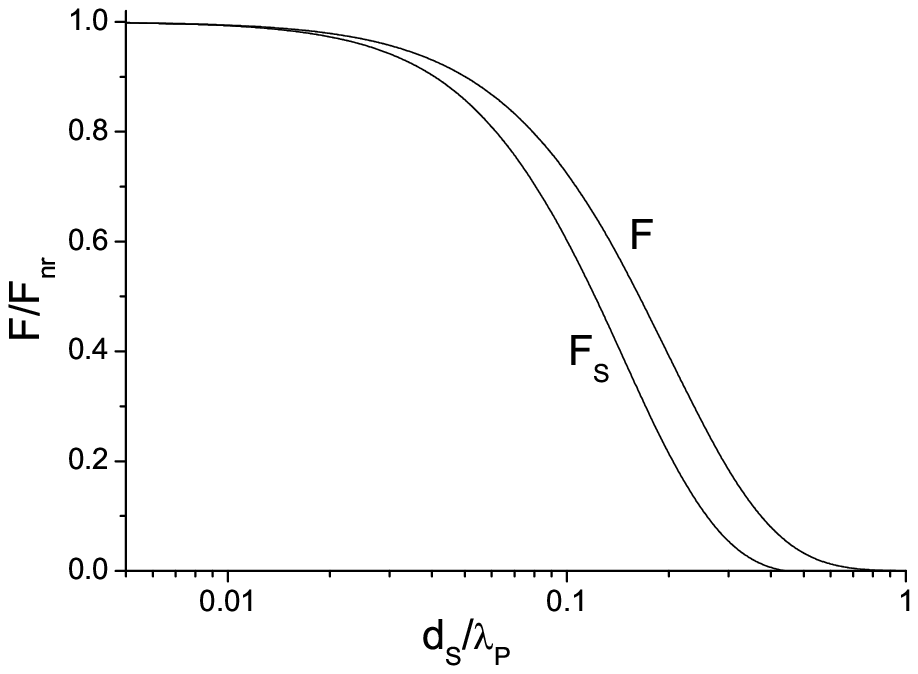}
\caption{\label{Fret}Vacuum-field pressure on a free standing
metal slab relative to its nonretarded value as a function of the
slab thickness. Lower line gives the surface polariton
contribution to the total vacuum-field pressure.}
\end{minipage}\hspace{1 cm}
\begin{minipage}{16pc}
\includegraphics[width=16pc]{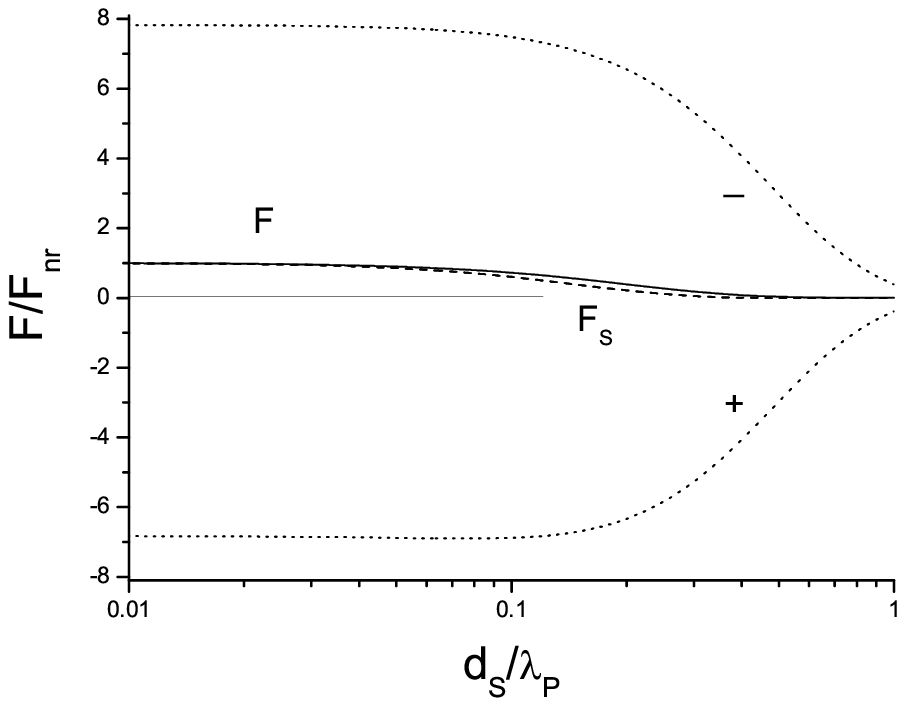}
\caption{\label{FSP}Surface polariton contribution (dashed line)
to the total vacuum-field pressure (full line) on a free standing
metal slab. Separate contributions $F^{\rm SP}_\pm$ of the two
surface polariton modes are presented by dotted lines}
\end{minipage}
\end{figure}

At intermediate slab thicknesses, the pressure must be calculated
using the exact result Eq. (\ref{ff}). The result of such a
calculation are presented in Figs. \ref{Fret} and \ref{FSP} where
we have plotted the pressure on a free standing slab relative to
its nonretarded value $F_{\rm nr}$ [Eq. (\ref{fnr})] and the
surface polariton contribution to it $F_S$, respectively. As seen,
owing to the field retardation, the true Casimir pressure $F$
deviates significantly from $F_{\rm nr}$ for slab thicknesses
$d_s/\lambda_P>0.01$ and can be in this region rather well
approximated by the contribution of two SP modes supported by the
slab $F_S$. Of course, at even larger slab thicknesses
($d_s/\lambda_P>1$), $F_S$ vanishes owing to the decoupling of SP
modes at two slab surfaces. The same happens to $F$ since, as
shown above, it attenuates exponentially for large thicknesses of
the slab.

\section{Metal slab between perfect mirrors}
Next we consider the other limiting case namely that of a metal
layer sandwiched between perfect mirrors. Such a situation is
described by Eqs. (\ref{fsg}) and (\ref{rq}) when letting
$R^q_i=\delta_{qp}-\delta_{qs}$ and $d_i=0$. From Eq. (\ref{rq})
it then follows that also $r^q_{s\pm}=\delta_{qp}-\delta_{qs}$ and
we have
\begin{equation}
F=-\frac{\hbar}{\pi^2}\int_0^\infty d\xi \int^\infty_0
dkk\kappa_s\frac{e^{-2\kappa_s d_s}}{1-e^{-2\kappa_s d_s}}.
\label{Fi}
\end{equation}
This formula can be significantly simplified by exploiting the
fact that the integral over $k$ vanishes when
$\xi\rightarrow\infty$. Following Schaden {\it et al}.
\cite{Schad}, we partially integrate over $\xi$ and simultaneously
introduce $\kappa_s$ as the integration variable in the integral
over $k$. This gives
\begin{eqnarray}
F&=&\frac{\hbar}{2\pi^2}\int_0^\infty d\xi\xi
\frac{d}{d\xi}\int^\infty_{\sqrt{\varepsilon_s(i\xi)}\xi/c}d\kappa_s\kappa^2_s
\sum_{q=p,s} \frac{e^{-2\kappa_sd_s}}
{1-e^{-2\kappa_sd_s}}\nonumber\\
&=&-\frac{\hbar}{\pi^2c^3}\int_0^\infty
d\xi\xi^3\sqrt{\varepsilon_s(i\xi)}
\frac{e^{-2\sqrt{\varepsilon_s(i\xi)}\xi d_s/c}}
{1-e^{-2\sqrt{\varepsilon_s(i\xi)}\xi d_s/c}},
\end{eqnarray}
where, in the second step, we have noted that for the free-electron
dielectric function $(d/d\xi)
\sqrt{\varepsilon_s(i\xi)}\xi=1/\sqrt{\varepsilon_s(i\xi)}$.
Finally, making the substitution
$y=\sqrt{\varepsilon_s(i\xi)}\xi/\omega_P$ and expanding the
integrand, we obtain for the pressure on the slab surfaces
\begin{equation}
\label{Fid} F=-\frac{\hbar
ck_P^4}{\pi^2}\sum_{n=1}^\infty\frac{d^2}{da_n^2}\frac{K_1(a_n)}{a_n},
\hspace{1 cm} a_n=2nk_Pd_s,
\end{equation}
where
\begin{equation}
K_1(a)=a\int_1^\infty dy\sqrt{y^2-1}e^{-ay}=\int_1^\infty
dy\frac{y}{\sqrt{y^2-1}}e^{-ay}
\end{equation}
is recognized as modified Bessel function of the second kind
\cite{GrRy,Math}.

To find pressure on a thin slab we use
\[\frac{d^2}{da^2}\frac{K_1(a)}{a}=\frac{6}{a^4}-\frac{1}{2a^2}+{\cal
O}(a^0),\hspace{1 cm} a\ll 1.\] This gives
\begin{equation}
f_s=F_C\left[1-\frac{5}{\pi^2}(k_Pd_s)^2+{\cal
O}[(k_Pd_s)^4]\right],\hspace{1 cm} F_C=-\frac{\pi^2\hbar
c}{240d_s^4},
\end{equation}
so that the pressure on a thin metal layer is, to the leading
order, given by famous Casimir result. Evidently, it is due to the
modes propagating through the layer. Indeed, as follows from Eq.
(\ref{DR}), this system supports only modes with frequencies
($\alpha_s=-in\pi/d_s$)
\[\omega^q_{n}(k)=c\sqrt{k_P^2+k^2+n^2\pi^2/d_s^2},\]
where $n$ is an integer. For $k_P\ll d_s^{-1}$, these mode
frequencies lead to the same Casimir energy as in the original
Casimir configuration. The pressure on a thick slab can be
obtained using
\[\frac{d^2}{da^2}\frac{K_1(a)}{a}=e^{-a}\sqrt{\frac{\pi}{2a^3}}
\left[1+\frac{27}{8a}+{\cal O}(a^{-2})\right],\hspace{1 cm} a\gg
1\] and is, to the leading order, again given by Eq. (\ref{fsf}).
These considerations are illustrated by the black full line in
Fig. \ref{Fcav} giving the pressure $F$ (relative to the Casimir
pressure $F_C$) on a free-electron metal layer between ideal
mirrors as a function of $k_Pd_s$.

\section{Metal slab in an ideal cavity}
\begin{figure}[htb]
\begin{minipage}{18pc}
 \includegraphics[width=18pc]{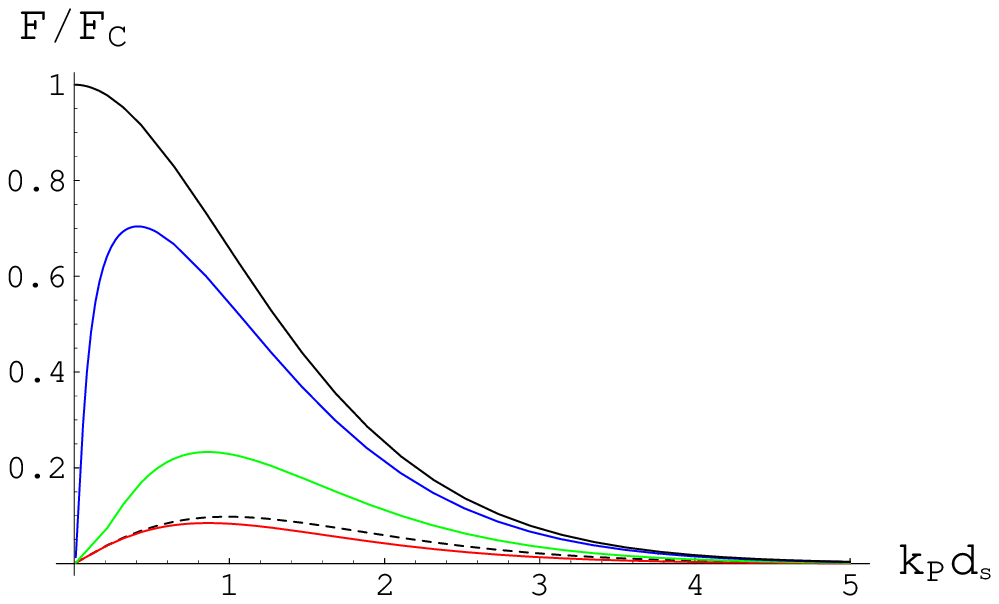}
 \caption{\label{Fcav}Relative vacuum-filed pressure on a metal slab in the center
 of an ideal cavity. The distance of the slab from the cavity mirrors
 is $k_pd=0$ (black full line), $k_pd=0.01$ (blue line), $k_pd=0.1$ (green line),
 $k_pd=1$ (red line) and $k_pd=\infty$ (black dashed line).}
\end{minipage}\hspace{2pc}
\begin{minipage}{18pc}
\includegraphics[width=18pc]{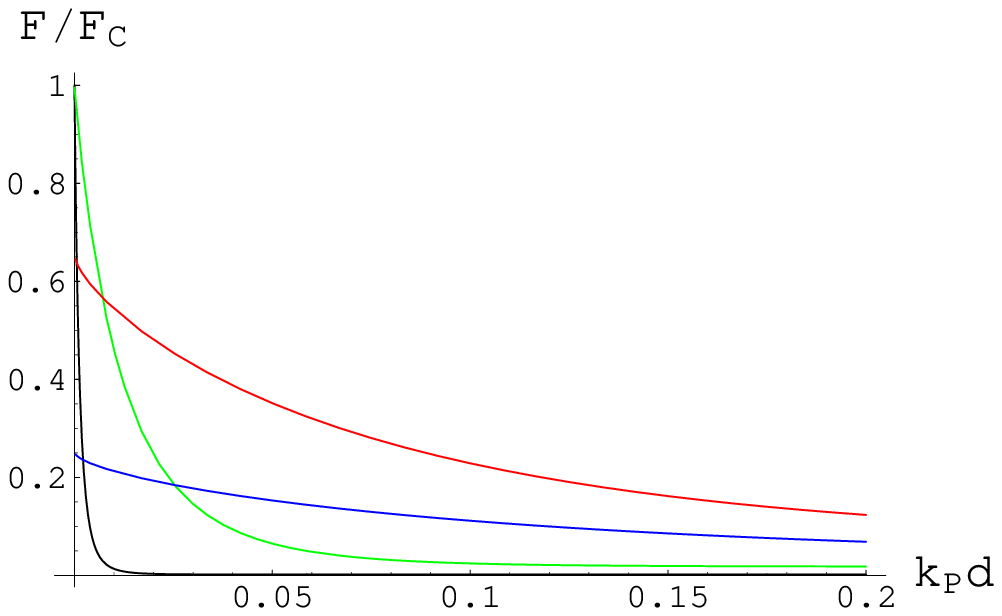}
 \caption{\label{Fc}Relative vacuum-filed pressure on a metal slab in the
 center of an ideal cavity as a function of its distance from
 cavity mirrors. The thickness of the slab is $k_pd_s=0.01$ (black line),
 $k_pd_s=0.1$ (green line), $k_pd_s=1$ (red line) and $k_pd_s=2$ (blue line).}
\end{minipage}
\end{figure}

The pressure on a metal slab in the center of a cavity with
perfectly reflecting mirrors is given by Eqs. (\ref{fsg}) and
(\ref{rq}) [or Eqs. (\ref{fs})] with
$R^q_i=\delta_{qp}-\delta_{qs}$ and $d_1=d_2\equiv d$ and can only
be calculated numerically. Its behaviour with the slab thickness
and the mirror-slab distance is illustrated in Figs. \ref{Fcav}
and \ref{Fc}.

\noindent As seen, cavity strongly affects the pressure on a thin
($k_Pd_s<0.1$) slab; one observes its strong drop from $F_C$ for
non zero mirror-slab distances. As shown above, with increasing
$d$ the pressure on a thin slab decreases to $F_{\rm nr}=0.19
k_Pd_s F_C$. We also recall that, whereas for $d=0$ the pressure
is due to modes propagating through the metal, at large
mirror-slab separations it is due solely to surface plasmon
(evanescent) modes. This can be understood when realizing that
nearby mirrors effectively damp surface modes since their field
cannot accommodate to the perfect mirror boundary condition.
Indeed, according to Eq. (\ref{DR}) (with $q=p$) the dispersion
relations of SP modes in the present configuration can be
rewritten as
\begin{equation}
\varepsilon(\omega)=-\frac{\alpha_s}{\alpha}\frac{\tanh^{\pm
1}\alpha_sd_s} {\tanh\alpha d}.
\end{equation}
Accordingly, when $d\rightarrow 0$ frequencies of SP modes tend to
zero. Thus when changing the cavity length, in addition to
changing its magnitude, one effectively modify the mode spectrum
of the pressure on the slab.

As expected, the effect of the cavity on the pressure on a thick
($k_Pd_s>1$) slab is less pronounced and diminish with the slab
thickness owing to the attenuation of the vacuum-field fluctuations
with $d_s$. We also observe from the red and black dashed curve in
Fig. \ref{Fcav} that for systems with mirror-slab distances
$k_Pd\geq 1$ the cavity effect on the pressure is very small and the
pressure on the slab differs very little (on this scale) from the
pressure on a free standing slab. Thus, the effect of the cavity on
the pressure on a slab in its center is largest for systems with the
mirror-slab distances $k_Pd<1$, as illustrated in Fig. \ref{Fc}.
Since $\lambda_P$ for (noble) metals is of the order of $10^2 {\rm
nm}$, this corresponds to the mirror-slab distances less than (say)
$20 {\rm nm}$.

\section{Summary}
In this work we have considered and performed a modal analysis of
the vacuum-field pressure on surfaces of a metal slab in the
center of a cavity adopting the plasma model for the metal and
assuming ideally reflecting cavity mirrors. We have confirmed
analytically previous results for the pressure on a thin and a
thick free-standing slab and derived an exact formula for the
pressure on a metallic layer between perfect mirrors. According to
these results, the pressure on a thin slab decreases with
increasing slab-mirror distances from the well-known Casimir
pressure $F_C$ to the quasistatic pressure $F_{\rm nr}=1.19
(d/\lambda_P)F_C$. In the first case it is entirely due to the
photonic modes propagating through the metal and in the second
case it is entirely due to the surface plasmon modes of the
free-standing slab. We recall that the pressure on metallic plates
in the standard Casimir configuration behaves with decreasing
their separation precisely in the same way: at large separations
between the plates it is equal to $F_C$ and is due to the photonic
modes \cite{Cas,Lif} whereas at small separations between the
plates it is equal to $F_{\rm nr}$ and is due to the surface
plasmon modes supported by the plates \cite{KNS,GILR,Henk,Bord}.
This similarity of the pressures in two configurations is lost in
the case of a thick slab as the pressure on its surfaces
exponentially decreases for large slab thicknesses. These
considerations demonstrate that a cavity may influence the
vacuum-field pressure on a thin metal slab (and its modal
structure) to a considerable extent.

\ack This work was supported by the Ministry of Science, Education
and Sport of the Republic of Croatia under contract No.
098-1191458-2870.

\section*{Referencess}

\end{document}